\documentclass[preprint,12pt]{elsarticle}

\usepackage{amssymb}

\journal{Nuclear Physics B}

\begin{document}

\begin{frontmatter}



\title{Neutrinos from Gamma Ray Bursts in the IceCube and ARA Era}


\author[label1,label2]{Dafne Guetta}

\address[label1]{ORT Braude College, Carmiel, Israel }
\address[label2]{INAF-OAR Monteporzio Catone, Italy}

\begin{abstract}
 In this review I discuss the  ultra-high energy neutrinos (UHEN) originated from Cosmic-Rays  propogation (GZK neutrinos) and from Gamma Ray Bursts (GRBs), and discuss their detectability in kilometers scale detectors like ARA and IceCube.  

While GZK neutrinos are expected from cosmic ray interactions on the CMB, the GRB neutrinos depend on the physics inside the sources. GRBs are predicted to emit UHEN in the prompt and  in the later 'after-glow' phase.

I discuss the constraints on the hadronic component of GRBs derived from the search of four years of IceCube data for a
prompt neutrino fux from gamma-ray bursts (GRBs) and more in general
I present the results of the search for high-energy neutrinos interacting within the IceCube detector between 2010 and
2013. 
\end{abstract}


\end{frontmatter}


\section{Introduction}
The detection of MeV  neutrinos emitted from the sun 
and from the supernovae allowed the understanding of the physics of these astrophysical objects
\cite{Bahcall04}, \cite{Raffelt05}. MeV neutrino ”telescopes”  are capable of detecting neutrinos from sources
close to our galaxy up to a distances $<$ 100 kp. 
 The main goal of the construction of high energy,$ >$ 1 TeV, neutrino
telescopes\cite{Gaisser95}  is the extension of the distance accessible to neutrino astronomy to cosmological
scales\cite{Waxman09}.

The existence of extra-Galactic high-energy neutrino sources is implied by cosmic-ray observations.
The cosmic-ray spectrum extends to energies $\sim 10^{20}$ eV, and is likely dominated
beyond $\sim 10^{19}$ eV by extra-Galactic sources.
 The origin of Cosmic Rays (CR) has been a tantalizing mystery ever since their discovery by Hess \cite{Hess} nearly a century ago. While “lower energy” CRs of up to $10^{16}$ eV are believed to originate from supernova explosions in our galaxy \cite{Blasi08}, the source of the more energetic CR whose energies can exceed $10^{19}$ eV remains unknown \cite{Cronin04}. 
Although ultrahigh energy cosmic rays (UHECR) are produced throughout the Universe, those observed at Earth must have been produced locally (within ~50 Mpc) since they lose energy while propagating through the Cosmic Microwave Background (CMB). This process (discussed by Greisen\cite{Greisen66} and Zatsepin and Kuzmin\cite{Zatsepin66} - "GZK") not only causes energy loss for the primary, but also creates secondary particles of extremely high energy \cite{Beresinsky69,Stecker73}, including neutrinos above $10^{17}$  eV.
The secondary particles, particularly Ultra High Energy Neutrinos (UHEN), can be used to explore the origins of UHECR\cite{Seckel05}. With no electric charge, neutrinos experience no scattering or energy loss, and so provide a probe of the source distribution even to high redshift.  Since UHECRs and the expected “GZK cutoff” have been observed, the expectation of a GZK neutrino flux is on very strong footing\cite{Mannheim98}. UHE neutrinos are also likely to be created at the acceleration sites of the UHECRs, pointing directly to the source from earth, and providing information on the role of hadrons in the acceleration.  

Only Active Galactic Nuclei\cite{Rachen93} and Gamma Ray Bursts, GRBs\cite{Milgrom95,Vietri95,Waxman95}, are believed to be capable of accelerating CRs to such enormous energies\cite{Dermer04}. 
Gamma Ray Bursts (GRBs) are powerful explosions, and are among the highest-redshift point sources observed. The most common phenomenological interpretation of these cosmological sources is through the so called fireball model \cite{fireball:2000, meszaAFTERGLOW, Meszaros06}. In this model, part of the energy is carried out (e.g., from a collapsed star) by hadrons at highly-relativistic energies, some of which
is dissipated internally and eventually reconverted into internal energy, which is then
radiated as $\gamma$-rays by synchrotron and inverse-Compton emission by shock-accelerated electrons.
As the fireball sweeps up ambient material, it energizes the surrounding medium through, e.g., forward shocks, which are believed to be responsible for the longer-wavelength afterglow emission \cite{meszaAFTERGLOW}.

If the GRB jet comprises PeV protons, it should produce energetic neutrinos through photon-hadron interactions.
The photons for this process can be supplied by the GRB gamma rays during its prompt phase, or during the afterglow phase \cite{WB97, Dermer02}.
These lead to the production of charged pions, which subsequently decay to produce neutrinos.
Within this picture, GRBs should produce neutrinos with energies of  $\sim 100$ TeV  from the
same region in which the GRB photons are produced \cite{dafneRATE}.
These neutrinos, if present, could be readily detected.
Hence, the detectability of TeV to PeV neutrinos depends on the presence of ($>$) PeV protons and on the efficiency at which their energy is converted into neutrinos, as compared to how much of the energy is in electrons, which is manifested primarily in the prompt GRB photon emission.

Neutrino astronomy has steadily progressed over the last half century, with successive generations of detectors achieving sensitivity to neutrinos with increasingly higher energies. With each increase in neutrino energy, the required detector increases in size to compensate for the dramatic decrease of the flux with energy.  
The high-energy neutrinos from GRBs and the GZK neutrinos discussed above should be detected by large neutrino telescopes, such as IceCube, the Askarian Radio array (ARA) and in the future KM3NeT \footnote{http://km3net.org/home.php}.
All these detectors  look for the Cherenkov radiation initiated by the neutrino interactions in the ice using either optical detection (IceCube,KM3Net) or Radio Frequency (RF) detection (ARA).

\section{Results from IceCube}
\begin{figure}[h]
  \centering
\includegraphics[width=0.7\textwidth]{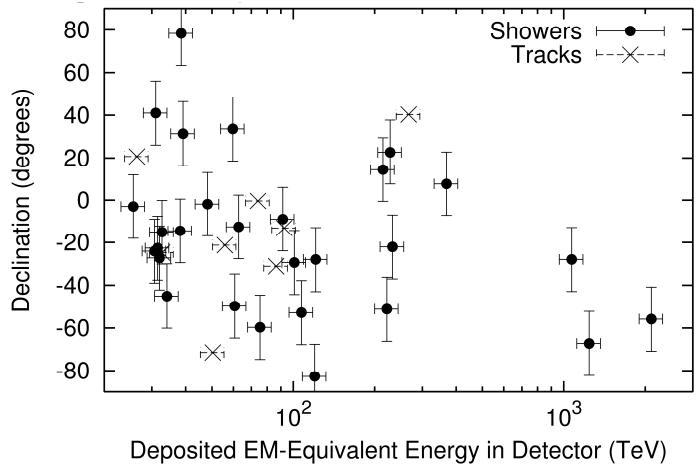}\\
  \caption{Arrival angles and deposited energies of the 37 events detected by Icecube.}
\label{NuNum}
\end{figure}

\begin{figure}[h]
  \centering
\includegraphics[width=0.7\textwidth]{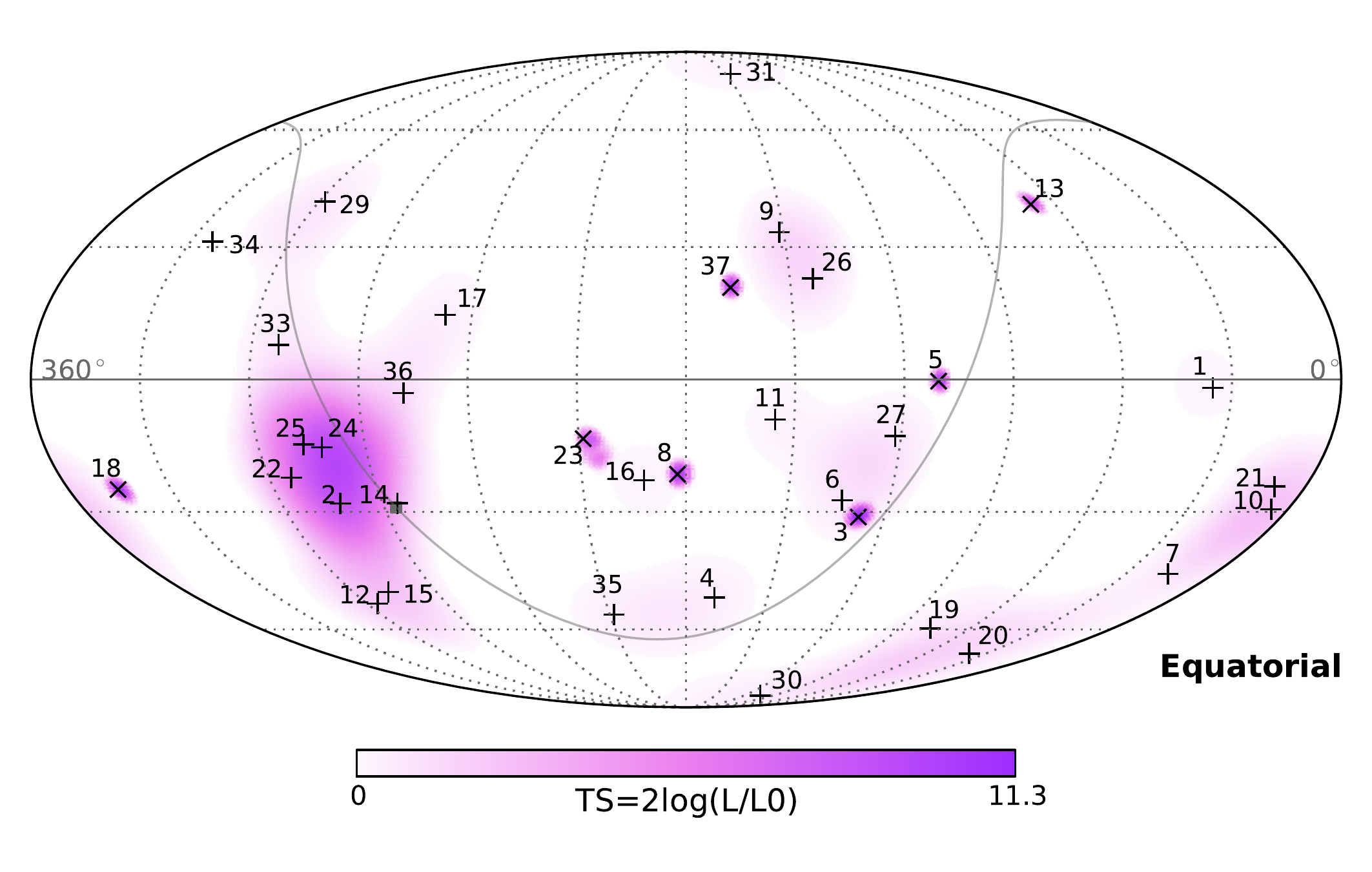}\\
  \caption{Arrival directions of the events in galactic coordinates.
Shower-like events (median angular resolution $\sim15^0$)
are marked with + and those containing muon tracks ($< 1^0$)
with x. The grey line denotes the
equatorial plane. Colors show the test statistic (TS) for the
point source clustering test at each location. No significant
clustering was observed.}
\label{NuNum}
\end{figure}
\
 IceCube, is a Cherenkov detector \cite{halzen2010} with photomultipliers (PMTs) at depths between 1450 and 2450 meters in the Antarctic ice designed specifically to detect neutrinos at TeV-PeV energies. Since May 2011 \cite{Aartsen}, IceCube has been working with a full capacity of 86 strings (IC86).
 IceCube analyses include a model-independent search for GRB neutrinos\cite{Abbasi2012}, and for other diffuse and point sources.
A search for high-energy neutrinos interacting within the IceCube detector between 2010 and
2013 provided the first evidence for a high-energy neutrino flux of extraterrestrial origin\cite{Icecube2014}. 
In the full 988-day sample, IceCube detected 37 events (Fig. 1) in the TeV-PeV range.
A purely atmospheric explanation for these events is
strongly disfavored by their properties \cite{IceCube3}.
The high galactic latitudes of many of the highest energy
events (Fig. 2) suggest at least some extragalactic component.
Moreover the intensity associated with the neutrino excess is much higher than that expected to originate
from interaction of cosmic-ray protons with interstellar gas in the Galaxy\cite{katz}
The flux, spectrum and angular distribution of the excess neutrino signal detected by IceCube between 50 TeV and 2 PeV suggest that the sources of these neutrinos are unlikely to be (unknown) Galactic sources.
If the sources were galactic the events were expected to be strongly concentrated along the galactic disk \cite{waxmanprod}. 
Results are consistent with an astrophysical flux in the 100 TeV - PeV
range at the level of $10^{-8}$GeV cm$^{-2}$s$^{-1}$sr$^{-1}$ per flavor. The data are consistent with expectations for equal fluxes
of all three neutrino flavors and with isotropic arrival directions, suggesting either numerous or
spatially extended sources. 
No evidence of neutrino emission from point-like or extended sources was found in four years
of IceCube data. Searches for emissions from point-like and extended sources anywhere in the
sky, from a pre-defined candidate source list and from stacked source catalogs all returned results
consistent with the background-only hypothesis. 90\% 
C.L. upper limits on the muon neutrino
fluxes for models from a variety of sources have been calculated and compared to predictions\cite{IceCube4}.
 The coincidence of the excess, $ E^{2}_{\nu}\phi_{\nu} = 3.6 \pm 1.2\,$$10^{-8}$ GeV/cm$^2$sr s, with the Waxman-Bahcall (WB) bound, $ E^{2}_{\nu}\phi_{WB} = 3.6 \pm 1.2 \,$$10^{-8}$ GeV/cm$^2$sr s, is probably a clue to the origin of IceCube’s
neutrinos. The most natural explanation of this coincidence is that both the neutrino excess
and the ultra-high energy, $> 10^{19}$ eV, cosmic-ray (UHECR) flux are produced by the same
population of cosmologically distributed sources  \cite{waxmanprod}.
The coincidence with the WB flux also support the hypothesis that the detected neutrinos have an extra-Galactic origin. The $\nu_{\mu}$:$\nu_e$:$\nu_{\tau}$ = 1 : 1 : 1 flavor ratio is consistent with that expected for neutrinos originating from pion decay in cosmologically distant sources, for which oscillations modify the
original 1 : 2 : 0 ratio to a 1 : 1 : 1 ratio \cite{learned}.
 In order to identify the sources of UHE neutrinos
a multiwavelength analysis is needed.

\section{Neutrinos from Gamma-Ray-Bursts}
\begin{figure}
  \centering
\includegraphics[width=0.7\textwidth]{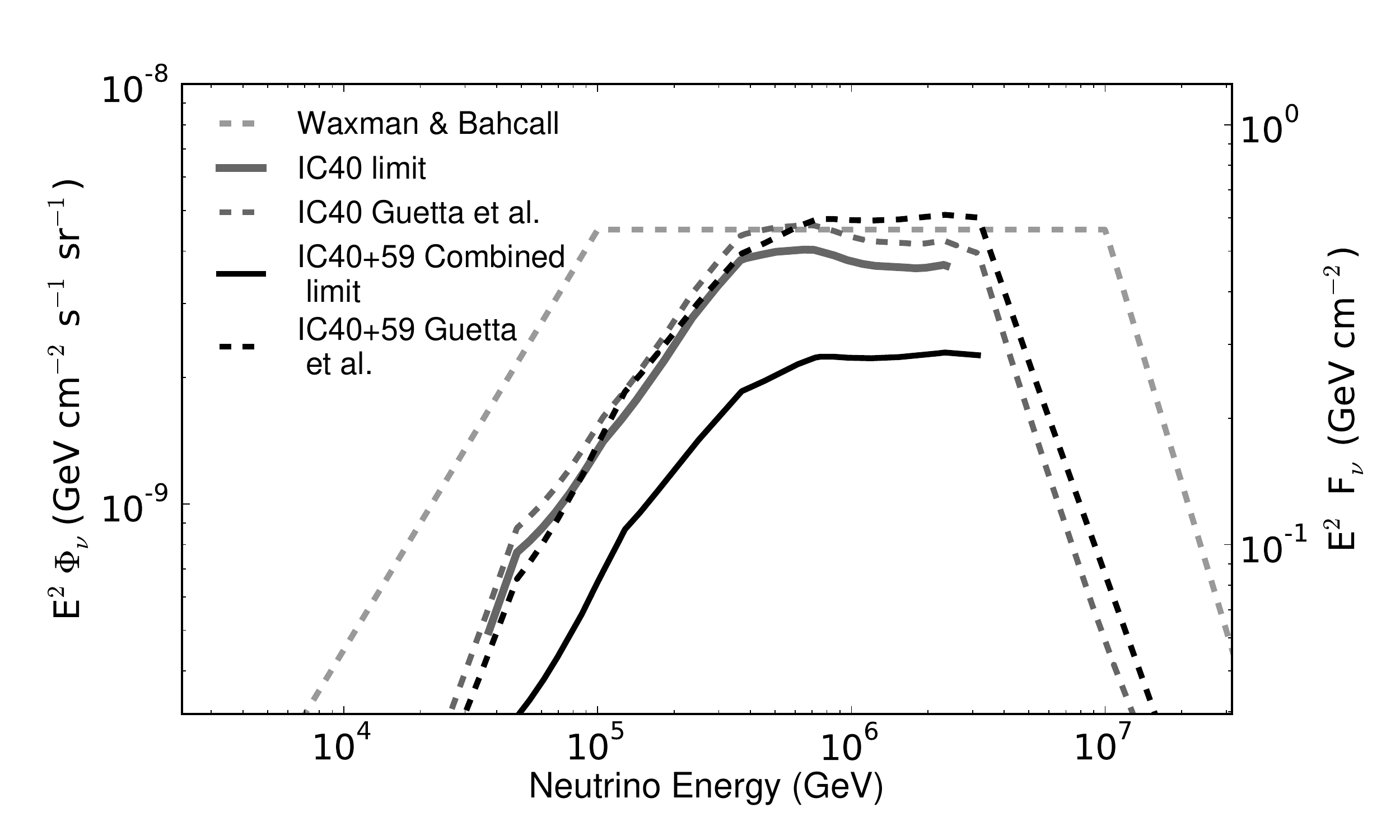}\\
  \caption{Comparison of results to predictions based on observed
gamma-ray spectra. The summed flux predictions
normalized to gamma-ray spectra\cite{meszabig,dafneRATE} is shown in dashed
lines; pay attention that the cosmic ray normalized Waxman-Bahcall flux \cite{WB97} shown 
here\cite{icecubenature} for reference is not correct and
should be a factor 10 smaller.}
\label{constraints}
\end{figure}
The high-energy neutrinos from GRBs should be detected by large neutrino telescopes, such as IceCube and in the future KM3NeT \footnote{http://km3net.org/home.php}.
Since GRB neutrino events need to be correlated both in time and in direction with the gamma-rays, they are sought after in small angular and short time windows.
Therefore the search for neutrino events should be coordinated with the gamma-ray telescopes that provide the GRB trigger.
In this context, IceCube has recently developed a powerful model-independent analysis tool for neutrinos detection, which is coincident in direction and in time to within 1,000 seconds with GRB flares reported by the gamma ray satellites.
IceCube reported no detection of any  GRB-associated neutrino in a data set taken from April 2008 to May 2010  \cite{icecubenature}; None of the high energy neutrinos reported for the next two years \cite{Aartsen_Sci} is GRB-associated either, and as far as we know no neutrino event has been associated with any GRB to date.
This non-detection is in conflict with earlier models \citep{WB97, meszabig,dafneRATE, otherbig, he2012},
all of which predicted the detection of approximately ten GRB neutrinos by IceCube during this period.
Those earlier estimates were largely calibrated based on the fireball hypothesis, and were motivated by the assumption that UHECRs are produced primarily by GRBs.
As shown in Fig.3 the IceCube results thus appear to rule out GRBs as  the main sources of UHECRs \citep{otherbig, icecubenature}.
This implies either that GRBs do not have the  ($>$)PeV protons, hypothesized in the fireball model,  or that the efficiency of neutrino production from these protons is much lower than what have been estimated \citep{small1,small2,small3}.
Recently the Icecube collaboration  set the most stringent limits yet on GRB neutrino
production using four years of IceCube data\cite{Icecube5}. 
They constrain parts of the parameter space relevant to the production of UHECRs in the
latest models. Because of the very low on-time and on-source background rate, the analysis grows
more sensitive almost linearly with time.

Recently Yacobi et al.2014\cite{yacobi} have used the data from the GRB Monitor (GBM) on board Fermi to calibrate the photon (representing electrons) energy content of the GRB jet.
Subsequently, they have compared this with the upper limit on proton (turned pion) energy content, given the non-detection of GRB neutrinos and derived a constraint on the fraction of energy that goes into hadrons to the one that goes into electrons, the ratio $f_\pi / f_e$.
\begin{figure}
  \centering
\includegraphics[width=0.7\textwidth]{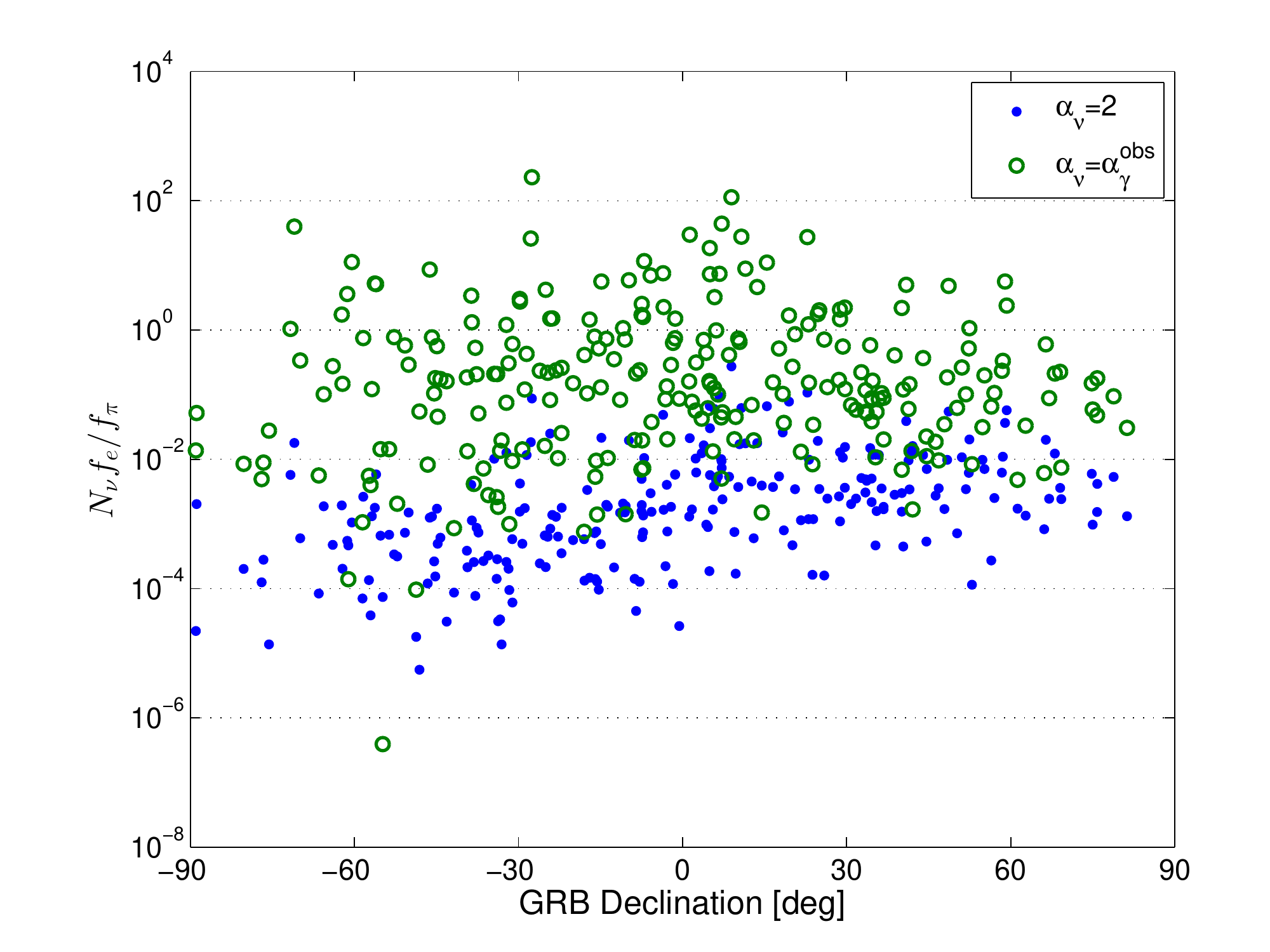}\\
  \caption{Expected number of $\nu _\mu$ from 250 GBM detected GRBs, which have a measured spectral slope, as a function of declination, and factored by the unknown electron to pion energy ratio $f_e / f_\pi$. The neutrino spectral slope $\alpha_{\nu}$ is assumed to be that of the photons
$\alpha_{\gamma}$. The results for $\alpha_{\nu}=2$ are reported for comparison\cite{yacobi}} \label{slope}
\end{figure}
The results for each individual GRB are plotted in Fig.4, which shows many GRBs yielding high values of $N_\nu (f_e / f_{\pi})$, and thus tightening the upper limit on the $f_\pi / f_e$  by more than two orders of magnitude.
Considering all the GBM sample Yacobi et al. 2014 found that the lack of detected neutrinos from Fermi/GBM GRBs since 2008 points to $f_\pi / f_e \lesssim 0.24$ with a 95\% CL.
The obtained value of $f_\pi / f_e \lesssim 0.2$ is still consistent with the values of $f_e\approx 1$ \citep{WB97}, and of $f_\pi \approx 0.2$ \citep{dafneRATE}.

Another constraint on the hadronic fractio of the GRB jet can be obtained using the data of the Large Area Telescope (LAT) on board Fermi.
The first catalog of LAT  includes 35 GRBs with gamma ray emission above 100  GeV \cite{Ackermann}.
Several models have been proposed to explain this high energy emission \cite{meszaAFTERGLOW, gg03, gpw11}  including hadronic models \cite{GupZhang, bd2000}.
The same photon-hadron process that produces the charged pions and subsequently the 100 TeV neutrinos, would also generate neutral pions that decay to photons of similar energy.
These high-energy photons have been hypothesized to cascade through pair production processes down to the GeV regime, where they can escape the jet and be observed by LAT.
Within this scenario, Yacobi et al 2014 use the observed GeV burst fluence to put another upper limit on the energy content of the protons in the jet.

The most conservative estimation for the hadronic contribution to the GeV photon fluence measured by LAT is $f_{\rm Had} \le 1$, i.e. all LAT fluence is hadronic (via  pair-photon cascades).
Using the typical ratio $F_{\rm GBM} / F_{\rm LAT} \approx 10$ \citep{Ackermann} 
Yacobi et al. 2014 constrain the typical GRB hadronic fraction to be $f_\pi / f_e \lesssim 0.3$.

Since the flux of neutrinos depends on the energy fraction of protons in the FB for a given burst energy, the flux or limit of the neutrino emission could constrain the hadronic component in the GRB jet and the hadronic emission models. Therefore a multi-wavelength analysis is mandatory in particular the synergy between the Swift, LAT, IceCube and ARA is fundamental to give constraints to the FB model and the hadronic emission models.

\section{GZK neutrinos and ARA}
Although ultrahigh energy cosmic rays (UHECR) likely originate
throughout the Universe, those observed at Earth
must be produced locally since such UHECR lose
energy while propagating through the CMB ($p\gamma -> n\pi^{+} -> n\mu + \nu_{\mu}$, etc.) producing GZK neutrinos.
The detection of UHE GZK neutrinos is an experimental challenge at the frontier of neutrino astronomy,
which has progressed over the last half century.
 
The flux of GZK neutrinos incident on earth ranges from $\sim0.01/{\rm km}^{2}/{\rm year}$ \cite{stanev08} to$\sim6/{\rm km}^{2}/{\rm year}$
 \cite{barger06},depending on the model considered.
This flux can be detected in dense, RF-transparent media such as ice via the Askaryan effect\cite{Askaryan62}.  There is approximately one UHE neutrino interaction occurring in each km$^3$ in the ice per year. The long RF attenuation length in the Antarctic ice allows a more efficient area coverage that makes it possible to construct detectors of order tens to hundreds of ${\rm km}^2$, and several small-scale pioneering efforts to develop this approach exist\cite{Rice,Aura,Anita}. Two years ago, a two-phased experiment named ARA (Askaryan Radio Array) was initiated, designed to ultimately accumulate hundreds of GZK neutrinos\cite{Ara}.  
The goal of Phase 1 of ARA (2010-2016)  is to make the first definitive observation of cosmogenic neutrinos;
Phase 2 of ARA (37 stations) 
would then accumulate the statistics necessary to carry out an expanded astrophysics and particle-physics science program.  The primary goal of ARA is to discover GZK neutrinos and to establish the spectrum.
  
The detection of  GZK neutrinos from ARA will  allow to understand the origin of 
the UHECR cutoff confirmed by the recent data of Auger and the  nature of the UHECR composition.

\begin{figure}
  \centering
\includegraphics[width=0.6 \textwidth]{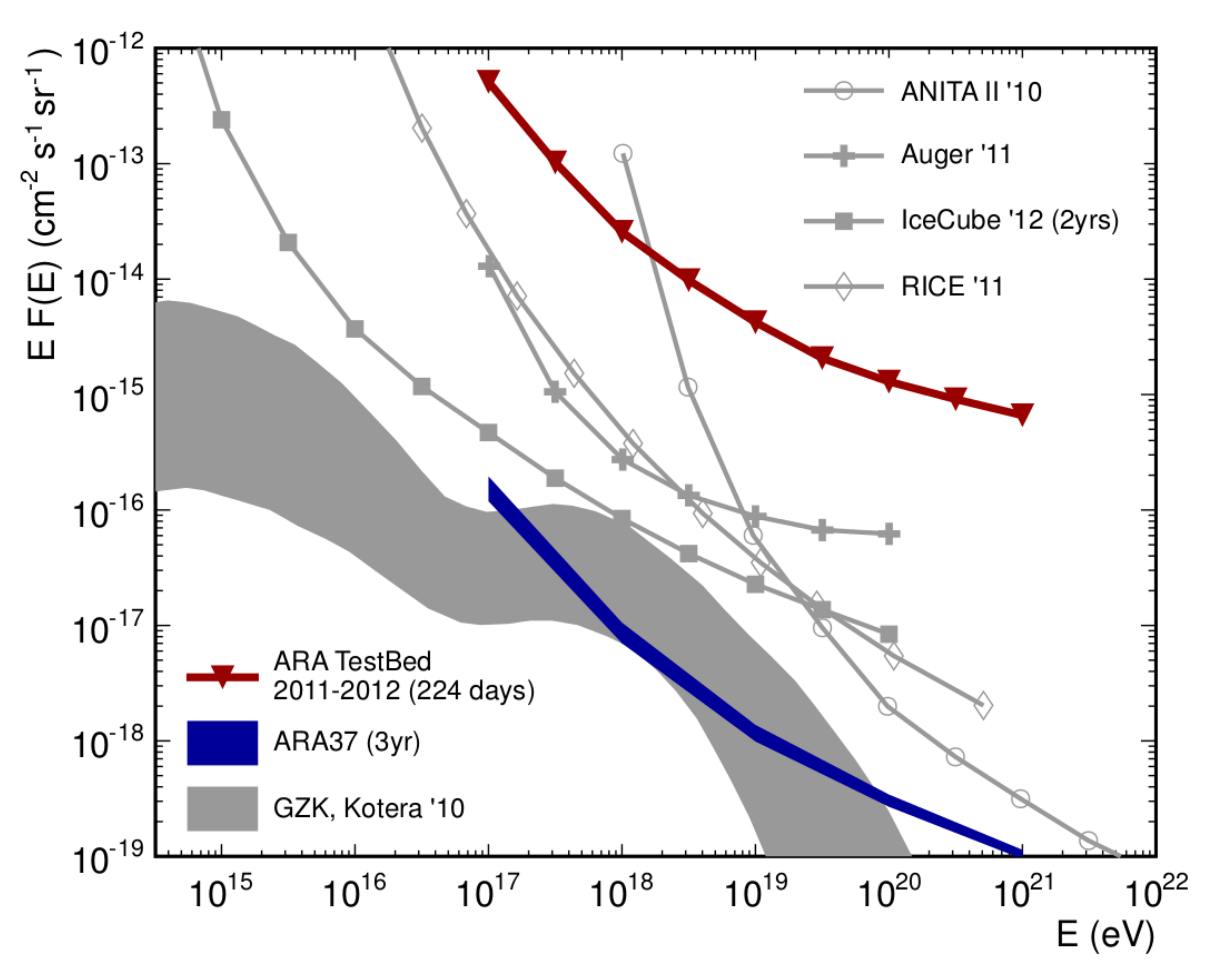}\\
  \caption{ The limits placed compared with the projected
ARA37 trigger-level sensitivity and results from
other experiments. } \label{slope}
\end{figure}

\section{Acknowledgements}
I Thank the Scientific Organization Committee for having invited me to give this plenary talk.
This research is supported by a grant from the U.S. Israel Binational Science Foundation.

\end{document}